# A Novel Boundary Matching Algorithm for Video Temporal Error Concealment

Seyed Mojtaba Marvasti-Zadeh
Department of Electrical and Computer Engineering, Yazd University
Yazd, Iran
E-mail: mojtaba.marvasti@stu.yazd.ac.ir

Hossein Ghanei-Yakhdan
Department of Electrical and Computer Engineering, Yazd University
Yazd, Iran
E-mail: hghaneiy@yazd.ac.ir

Shohreh Kasaei
Department of Computer Engineering, Sharif University of Technology
Tehran, Iran
E-mail: skasaei@sharif.edu

*Abstract*— With the fast growth of communication networks, the video data transmission from these networks is extremely vulnerable. Error concealment is a technique to estimate the damaged data by employing the correctly received data at the decoder. In this paper, an efficient boundary matching algorithm for estimating damaged *motion vectors* (MVs) is proposed. The proposed algorithm performs error concealment for each damaged *macroblock* (MB) according to the list of identified priority of each frame. It then uses a classic boundary matching criterion or the proposed boundary matching criterion adaptively to identify matching distortion in each boundary of candidate MB. Finally, the candidate MV with minimum distortion is selected as an MV of damaged MB and the list of priorities is updated. Experimental results show that the proposed algorithm improves both objective and subjective qualities of reconstructed frames without any significant increase in computational cost. The PSNR for test sequences in some frames is increased about 4.7, 4.5, and 4.4 dB compared to the classic boundary matching, directional boundary matching, and directional temporal boundary matching algorithm, respectively.

*Index Terms*— Temporal error concealment, Motion vector estimation, Boundary matching algorithm.

## I. Introduction

Video data compression is a process to decrease the digital data rate by removing redundant data. Elimination of video data redundancy may cause sensitivity to channel errors. In order to conceal these errors and enhancement of the reconstructed frames visual quality, it is used the error concealment techniques. These techniques are classified according to the used video sequences properties. In general, error concealment techniques are divided into three main categories: spatial domain [1-3], frequency domain [4-6], and temporal domain [7-34].

Spatial error concealment techniques use spatial redundancies among frame pixels to recover the damaged MB. Frequency error concealment techniques use data of adjacent MBs in the frequency domain for error concealment. Temporal error concealment exploits temporal redundancy among consecutive frames for error concealing of damaged MB. If there are no scene changes in consecutive frames, most objects of the current frame can be found in the previous frame. In this paper, we have focused on temporal error concealment.

The simplest temporal error concealment method is the *temporal replacement* (TR) [7]. In that method, all damaged MVs are replaced by zeros. The method is useful when there is a low motion among consecutive frames. In [8], a whole frame loss error concealment algorithm is proposed to further refine the TR. There are also some other simple methods which have special usages. Some of these methods [9] use the corresponding MB's MV from the previous frame which has a better performance, assuming that the motion in video sequences is smooth. Another simple and common method is the use of the average or median of adjacent MVs of damaged MB [10]. The simulation results show that the use of median is better than average method. In [11], five simple error concealment methods are analyzed with two similarity metrics.

In the MV interpolation method [12], the MV of each 4×4 block is estimated by interpolation from MVs of adjacent MBs. The distance between adjacent blocks is used as their weights.

The *Lagrange interpolation* (LI) method [13], is also a simple and useful method for MV recovering of 4×4 blocks. It supposes that the damaged MVs are in the range of MVs from adjacent MBs. Obviously, if this assumption is invalid, the results will not be satisfactory.

The classic *boundary matching algorithm* (BMA) [14] uses the smoothness assumption from boundary pixels of damaged MB. The algorithm recovers the damaged MB by minimizing the boundary matching distortion among inner and outer boundaries of the reconstructed MB. The BMA method achieves very well results in estimating the MV of the damaged MB. However, slanting edges and rapid gray-level changes may cause extensive variations, which in turn decreases the BMA performance.

*Gao and Lie* [15] proposed a post-processing method for BMA with the use of Kalman filter. At first, the damaged MV is estimated by BMA. However, because of less information from boundary pixels, most of the estimated MVs are not accurate. Therefore, the obtained MVs are corrected by using the Kalman filter.

The *edge adaptive boundary matching algorithm* (EA-BMA) [16] uses a mask proportional to the edge strength in damaged blocks' neighbors. The BMA-based algorithm employs the masks outside a damaged block instead of predicting the boundary pixels of the damaged block. Finally, the best-selected MV from matching process is used for damaged block.

An effective temporal error concealment algorithm [17], constructs a limited candidate MV set among the MVs of neighboring MBs and extrapolates MVs. It selects the best MV from the limited candidate MV set by using the BMA to conceal the corrupted MB.

*Choi and Jeon* [18] proposed an error concealment technique with block boundary smoothing to improve the video subjective quality. It uses the weighted boundary pixels of reference block to decrease the blocking artifacts compared to conventional temporal error concealment methods.

*Huang and Lien* [19] proposed a temporal error concealment technique using a self-organizing map. They used a self-organizing map as a predictor to estimate the MVs of damaged MBs. The estimated MVs were utilized to reconstruct the damaged MB by exploiting the spatial information from reference frames via employing a boundary matching criterion. The dynamic temporal error concealment technique using a *competitive neural network* (CNN) [20] uses a CNN predictor or BMA method for estimating damaged MB's MV. Different methods are performed based on the video scene motion.

In a fuzzy reasoning-based temporal error concealment method [21], two measuring criterions, namely *side match distortion* (SMD) and *sum of absolute difference* (SAD), are considered together for estimating damaged MVs. Thus, the method is adopted to balance the effects of SMD and SAD to accomplish the judgment more accurately for candidate MVs. Also in [22], a fuzzy metric based on Sugeno fuzzy integral is used as the criterion to compare the candidate MVs. Unlike conventional metrics, this metric is more compatible with *human visual system* (HVS).

*Araghi et al.* [23] proposed a method for MV optimization of damaged MB with the two best MVs, which have obtained from BMA. Furthermore, a pre-processing step for determination of a proper MVs set is presented. According to BMA criterion with reliability coefficient in double weighted MVs algorithm [24], the two best MVs are selected. Then, the optimal MV is calculated by weighting the MVs in terms of their accuracy.

*Thaipanich et al.* [25] proposed an *outer boundary matching algorithm* (OBMA) to estimate damaged MVs. The OBMA uses spatial and temporal smoothness of damaged MB boundaries to conceal the damaged MB. This algorithm with minimizing the boundary matching distortion among outer boundaries of damaged MB and outer boundaries of reconstructed MB, recovers the damaged MB. Also in [26], the TR and the improved outer boundary matching algorithm have used for dynamical error concealment in inter-frames of videos.

In *spatio-temporal boundary matching algorithm* (STBMA) [27] has been used from BMA and OBMA criterions and side smoothness criterion of damaged MB for estimating the damaged MV.

The MVs interpolation method [28] interpolates the MVs of lost blocks in current frame using extrapolated MVs from the previous frame. It increases the accuracy in recovering the corrupted block MVs.

*Wu et al.* [29] proposed an enhanced edge-sensitive processing order for temporal error concealment algorithm. It uses an efficient processing order for error concealment by considering the side information of neighboring blocks. In addition, a MV searching algorithm for determining the best MV is presented.

An adaptive error concealment mechanism with the use of decision tree is proposed in [30] for damaged MB error concealment. It uses different spatial and temporal error concealment methods in terms of spatial and temporal features of the video sequences.

*Wang et al.* [31] proposed an integrated temporal error concealment technique for H.264/AVC. It switches between two modes, adaptively. The first mode is a conventional temporal error concealment step. The second mode is an integrated mode that obtains by integrating two temporal error concealment approaches with an adaptive weight. It can obtain the optimal recovery data for damaged MBs.

*Wu et al.* [32] proposed a spatial-temporal error concealment algorithm for H.264/AVC. In this algorithm, a frame-level scene-change detection is applied. If the scene change occurs the spatial error concealment is applied, otherwise the temporal error concealment is applied. For temporal error concealment, a prediction-based motion vector estimation scheme is applied to obtain the final MV.

Most of the conventional error concealment methods (such as BMA) apply only one direction to calculate the differences among boundaries in the boundary distortion function. The *directional boundary matching* (DBM) method [33] determines the direction of comparison for each boundary pixel of the candidate MB. Then, every boundary pixel in determined direction is compared with a pixel of the outer boundary of the damaged MB. It tries to improve the accuracy of damaged MV estimation.

The *directional temporal boundary matching algorithm* (DTBMA) [34] estimates the real boundary direction by considering pixel differences in three directions for temporal error concealment. If candidate MBs are from previous stages of error concealment, it will not lead to satisfactory results. Although, in general, it has a better performance compare to BMA and DBM.

To solve the problems of conventional algorithms, an efficient boundary matching algorithm is proposed to estimate the damaged MVs more accurately. In this algorithm, the damaged MBs in each frame are reconstructed according to a priority list of identified error concealment. Then, this list is updated after reconstruction of each damaged MB. Moreover, this algorithm uses the classic boundary matching criterion or the proposed boundary matching criterion to identify boundary matching distortion for each candidate MBs' boundary, adaptively. Therefore, the proposed algorithm obtains more accurate estimation of damaged MV. It also prevents from error propagation in next frames.

The rest of this paper is organized as follows. First in Section II, the BMA is described briefly. Afterwards, Section III presents the proposed algorithm in detail. In Section IV, the experimental results are shown and finally Section V, concludes the paper.

## II. BOUNDARY MATCHING ALGORITHM

The algorithm assumes that there are high spatial correlations in undamaged image pixels. In the classic *boundary matching criterion* (BMC) the inner boundary pixels of candidate MB are compared with the outer boundary pixels of damaged MB in the current frame. Outer boundaries of damaged MB and inner boundaries of candidate MB are shown in Fig. 1.

The classic boundary matching criterion is defined for each boundary of damaged MB using

$$BMC_{top} = \sum_{n=0}^{S-1} |f_{cur}(i+n, j-1) - f_{ref}(i+vi+n, j+vj)| \quad (1)$$

$$BMC_{bottom} = \sum_{n=0}^{S-1} |f_{cur}(i+n, j+S) - f_{ref}(i+vi+n, j+vj+S-1)| \quad (2)$$

$$BMC_{left} = \sum_{n=0}^{S-1} |f_{cur}(i-1, j+n) - f_{ref}(i+vi, j+vj+n)| \quad (3)$$

$$BMC_{right} = \sum_{n=0}^{S-1} |f_{cur}(i+S, j+n) - f_{ref}(i+vi+S-1, j+vj+n)| \quad (4)$$

where $(i,j)$ denotes the coordinate of top-left pixel in the damaged MB, $f_{cur}(.,.)$ is the indicator of current frame, $f_{ref}(.,.)$ is the indicator of reference frame, $MV(vi,vj)$ is the candidate MV, and $S$ is the number of available pixels in each boundary. Also, *top*, *bottom*, *left*, and *right* are referred to top, left, bottom, and right boundaries of the damaged MB.

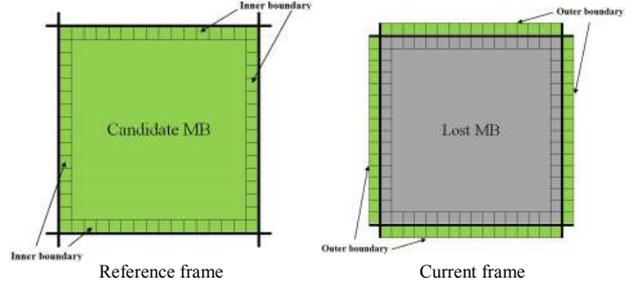

Figure 1. Illustration of boundaries in boundary matching criterion.

If one of outer boundaries of the damaged MB is unavailable, the corresponding part with the boundary from equation (1) to (4) is also unavailable. Finally, the candidate MV that leads to the minimum value of the classic matching distortion function ($BMC_{total}$) is selected as the MV of the damaged MB using

$$BMC_{total} = BMC_{top} + BMC_{bottom} + BMC_{left} + BMC_{right} \quad (5)$$

## III. PROPOSED ALGORITHM

In order to focus on the problem of concealing the damaged MBs, it is assumed that the positions of damaged MBs in video frames have been known. Some of the error detection methods at decoder are reviewed in [35].

According to the architecture of the proposed algorithm (Fig. 2), the numbers of correct spatial adjacent MBs for each damaged MB are identified after calculating the coordinate of damaged MB in current frame. Using these numbers, the list of error concealment priority from the damaged MB in current frame is identified. In this list, a damaged MB with more priority has more current boundaries.

Using this list in each stage, the MB with the highest priority is selected for error concealment. The list is updated after error concealment of each damaged MB. The updating includes deletion of concealed MB and the number of its neighbors and addition of a neighbor of the spatial adjacent MBs if they are damaged.

After that, one of the damaged MBs with the highest priority is selected for error concealment. The MVs of top, bottom, left, and right MBs from the damaged MB and the MV corresponding to the selected damaged MV from the previous frame are exploited.

The adapted candidate MV set includes the adjacent MVs of damaged MB, the mean and median MVs of them, the zero MV, and the corresponding MV of the damaged MV of the reference frame. In order to improve the accuracy of the proposed algorithm, the adjacent damaged MVs will be eliminated from the candidate MV set.

The proposed boundary matching algorithm uses two boundary matching criterions to compare the candidate MB boundaries and the adjacent boundaries of the damaged MB, adaptively. The first criterion is the classic boundary matching criterion described in Section II. The second one is the proposed boundary matching criterion described in the following.

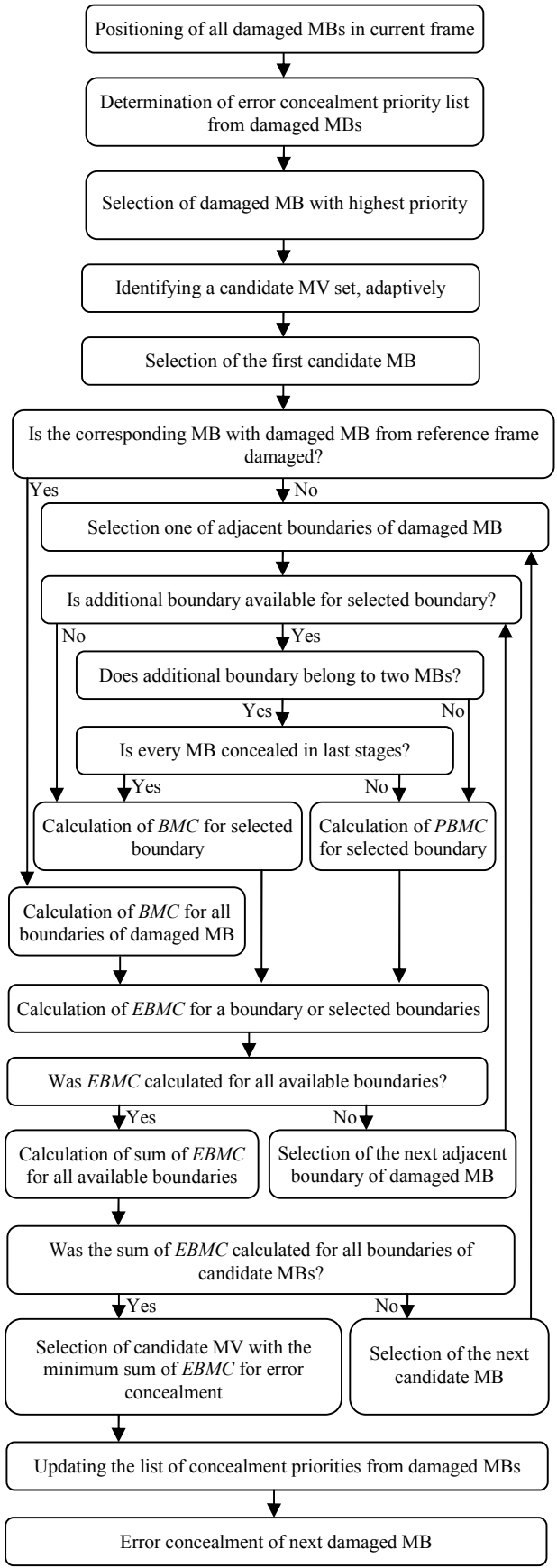

Figure 2. Block diagram of proposed algorithm.

If adjacent MBs of damaged MB are available, their positions are identified in the reference frame. Hence, the proposed boundary matching criterion exploits from the reference frame, the outer boundaries of the MBs which are overlapped with inner boundaries of damaged MB; these boundaries are called "additional boundaries". Additional boundaries for each damaged MB (if they are available), include the bottom outer boundary of top adjacent MB, the top outer boundary of bottom adjacent MB, the right outer boundary of left adjacent MB, and the left outer boundary of right adjacent MB.

If the spatial corresponding MB with the damaged MB is concealed in the reference frame (it results from the last error concealment stages), the proposed algorithm uses just the classic boundary matching criterion for error concealment. Furthermore, if each of additional boundaries is belonged to the different MBs in the reference frame and each of these MBs is concealed, the additional boundary is ignored and the classic boundary matching criterion is used for comparing that boundary. Otherwise, the *proposed boundary matching criterion* (PBMC) is defined using

$$PBMC_{top} = \sum_{n=0}^{S-1} | f_{ref}(i+vi+n, j+vj) - f_{ref}(i+vtx+n, j+vty) | \qquad (6)$$

$$PBMC_{bottom} = \sum_{n=0}^{S-1} | f_{ref}(i+vi+n, j+vj+S-1) - f_{ref}(i+vbx+n, j+vby+S-1) | \qquad (7)$$

$$PBMC_{left} = \sum_{n=0}^{S-1} | f_{ref}(i+vi, j+vj+n) - f_{ref}(i+vlx, j+vly+n) | \qquad (8)$$

$$PBMC_{right} = \sum_{n=0}^{S-1} | f_{ref}(i+vi+S-1, j+vj+n) - f_{ref}(i+vrx+S-1, j+vry+n) | \qquad (9)$$

where $MV(vtx,vty)$, $MV(vbx,vby)$, $MV(vlx,vly)$, and $MV(vrx,vry)$ denote the MVs of adjacent MBs from top, bottom, left, and right damaged MBs, respectively; if they are available. Also, $(v.x,v.y)$ are the components of MVs from adjacent MBs in $x$ and $y$ sides. Also, $t$, $b$, $l$, and $r$ refer to top, bottom, left, and right adjacent MBs of damaged MB, respectively.

Finally, according to (10) the candidate MV is selected as the MV of damaged MB with minimum boundary distortion of the proposed $EBMC_{position}$ in all boundaries

$$EBMC_{position} = \underset{position \in \{top,left,bottom,right\}}{\arg\min} \langle BMC_{position}, PBMC_{position} \rangle \qquad (10)$$

where *position* is referred to the position of each MV from damaged MB.

## IV. EXPERIMENTAL RESULTS

In order to evaluate the performance of the proposed algorithm, various types of CIF (325×288) video test sequences including "Bus", "Mother and Daughter", and "Foreman" and also QCIF (176×144) video test sequences including "Walk", "Mother and Daughter", and "Miss America" were used. These test sequences have different motion types in consecutive frames.

Video test sequences are encoded in 4:2:0 format. The size of the MBs is 16×16. To calculate MVs, the block matching algorithm [36] with exhaustive search (full search) and the search parameter of seven (p=7) is used. Error in video frames with MB missing rate of 5%, 10%, and 20% in each frame are created, randomly.

To evaluate the performance of the proposed algorithm, the damaged MVs are recovered by different algorithms. The performance of the proposed algorithm is compared with BMA, DBM, and DTBMA methods. For increasing the validity of the tests, the experiments on each algorithm are performed 20 times and the average values are used as the final result.

Fig. 3 shows the average PSNR values for reconstructed CIF test sequences of "Bus", "Mother and Daughter", and "Foreman", respectively. Also, Fig. 4 shows the average PSNR values for the reconstructed QCIF video test sequences of "Mother and Daughter", "Walk", and "Miss America, respectively.

Experimental results shown in Fig. 3 indicate that the proposed algorithm increases the PSNR for CIF video test sequences in some frames about 2.93, 2.96, and 2.82 dB compared to BMA, DBM, and DTBMA methods, respectively. Also, according to Fig. 4, the proposed algorithm increases the PSNR for QCIF test sequences in some frames about 4.71, 4.58, and 4.46 dB compared to BMA, DBM, and DTBMA methods, respectively.

The average PSNR values (in dB) of luminance for 30 frames of CIF sequences and 60 frames of QCIF sequences are listed in Table I. To compare the quality and time complexity of various error concealment algorithms the same PC (Intel Core i5, 2.4 GHz) is used. The average reconstruction time (in msec) for 30 frames of CIF sequences and 60 frames of QCIF sequences with average MB missing rate of 10% is listed in Table II.

According to Table I, the proposed algorithm yields higher average PSNR performance than BMA, DBM, and DTBMA in both CIF and QCIF resolutions. From Table I we can observe that the proposed algorithm in CIF resolution can improve the average PSNR performance by up to 1.8727, 1.9257, and 1.8116 dB, compared with BMA, DBM, and DTBMA, respectively. Also, the proposed algorithm in QCIF resolution can improve the average PSNR performance by up to 1.5312, 1.4935, and 1.2577 dB, compared with BMA, DBM, and DTBMA, respectively. Also, from Tables I and II we can observe that the proposed algorithm estimates the MVs of the damaged MBs without considerable increase in computational complexity compared to BMA, DBM, and DTBMA.

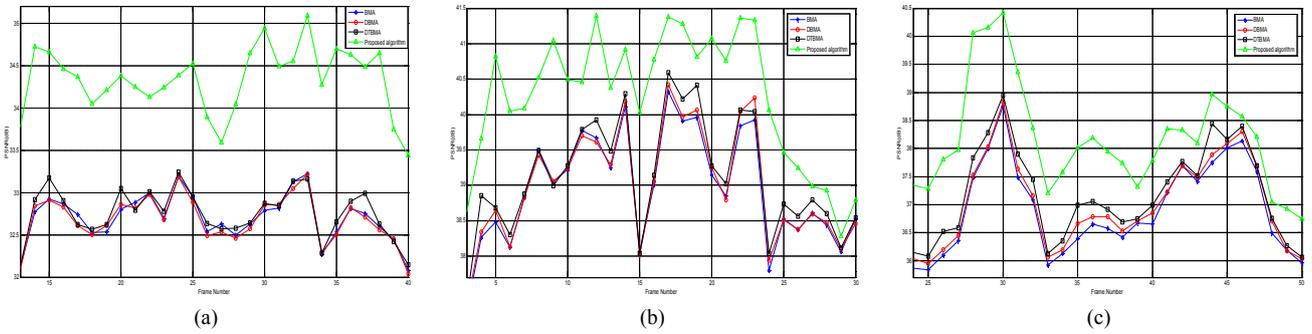

Figure 3. PSNR values of different CIF video test sequences with average MB missing rate of 20%: (a) "Bus"  (b) "Mother and Daughter"  (c) "Foreman".

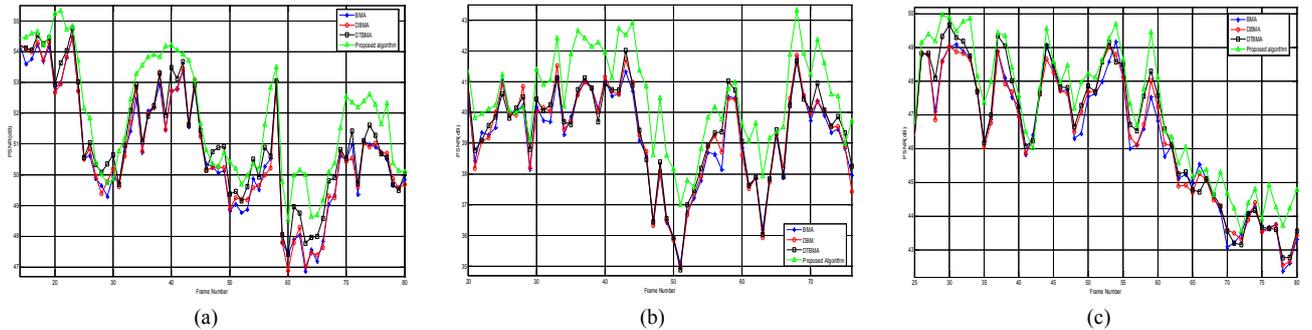

Figure 4. PSNR values of different QCIF video test sequences with average MB missing rate of 20%: (a) "Mother and Daughter"  (b) "Walk"  (c) "Miss America".

TABLE I. AVERAGE PSNR FOR VIDEO TEST SEQUENCES WITH DIFFERENT ERROR CONCEALMENT ALGORITHMS.

| Video Resolution | Video Sequence | Error Concealment Algorithm | Average MB Missing Rate | | |
|---|---|---|---|---|---|
| | | | *5%* | *10%* | *20%* |
| CIF | Bus | BMA | 38.1736 | 34.8043 | 31.9832 |
| | | DBM | 38.1423 | 34.7514 | 31.9656 |
| | | DTBMA | 38.3976 | 34.8655 | 32.0349 |
| | | Proposed Algorithm | 39.5173 | 36.6770 | 33.5487 |
| | Mother and Daughter | BMA | 43.3521 | 40.5038 | 37.8773 |
| | | DBM | 43.4700 | 40.5407 | 37.9161 |
| | | DTBMA | 43.5226 | 40.6360 | 38.0256 |
| | | Proposed Algorithm | 44.4984 | 41.9167 | 39.1109 |
| | Foreman | BMA | 41.4184 | 38.7466 | 35.6586 |
| | | DBM | 41.5632 | 38.8565 | 35.7570 |
| | | DTBMA | 41.8779 | 39.0506 | 35.9118 |
| | | Proposed Algorithm | 42.4773 | 39.9759 | 36.8464 |
| QCIF | Mother and Daughter | BMA | 60.3064 | 54.6166 | 50.0784 |
| | | DBM | 60.3441 | 54.6706 | 50.1183 |
| | | DTBMA | 60.7757 | 54.9260 | 50.4305 |
| | | Proposed Algorithm | 61.8376 | 55.5045 | 51.1868 |
| | Walk | BMA | 49.2718 | 43.6730 | 38.7396 |
| | | DBM | 49.2980 | 43.7340 | 38.8072 |
| | | DTBMA | 49.3812 | 43.8938 | 38.9161 |
| | | Proposed Algorithm | 50.5197 | 45.1515 | 39.9987 |
| | Miss America | BMA | 56.6534 | 50.2796 | 45.8118 |
| | | DBM | 56.5755 | 50.2537 | 45.8281 |
| | | DTBMA | 56.6606 | 50.4797 | 46.0167 |
| | | Proposed Algorithm | 56.9835 | 50.8658 | 46.5073 |

TABLE II. AVERAGE RECONSTRUCTION TIME PER MB FOR AVERAGE MB MISSING RATE OF 10%.

| Video Resolution | Video Sequence | Error Concealment Algorithm | | | |
|---|---|---|---|---|---|
| | | *BMA* | *DBM* | *DTBMA* | *Proposed Algorithm* |
| CIF | Bus | 2.7993 | 3.3019 | 3.3344 | 3.1977 |
| | Mother and Daughter | 2.7869 | 3.3274 | 3.3723 | 3.2179 |
| | Foreman | 2.6302 | 3.1539 | 3.1872 | 3.0375 |
| QCIF | Mother and Daughter | 4.9604 | 5.5570 | 5.5986 | 5.2034 |
| | Walk | 3.8730 | 4.4414 | 4.5425 | 4.3046 |
| | Miss America | 3.7974 | 4.3264 | 4.3615 | 4.0332 |

Fig. 5(a) shows an error-free frame in CIF test sequence "Bus" (frame 16). Fig. 5(b) shows the damaged frame with average MB missing rate of 20%. The reconstructed frames using the BMA, DBM, DTBMA, and the proposed algorithm are shown in Figs. 5(c), 5(d), 5(e), and 5(f), respectively. Also, Fig. 6(a) shows an error-free frame in QCIF test sequence "Walk" (frame 31). Fig. 6(b) shows the damaged frame with average MB missing rate of 20%. Reconstructed frames using the BMA, DBM, DTBMA, and the proposed algorithm are shown in Figs. 6(c), 6(d), 6(e), and 6(f), respectively. According to Figs. 5 and 6, the proposed algorithm improves the subjective quality of reconstructed video frames.

The BMA assumed that image pixels in original image have a high spatial correlation. If the boundary pixels of damaged MB are smooth, this algorithm is acceptable; otherwise, the results are not preferable. In addition, this algorithm can correctly estimate the MV of damaged MB when the damaged MB is on vertical and horizontal boundaries. However, it cannot be useful for oblique edges and cannot estimate the damaged MV well.

In DBM, the direction of comparison in each boundary pixel is identified from two inner boundaries of the candidate MB. However, in DTBMA this direction is identified by inner and outer boundaries of the candidate MB. When the direction of edges on boundaries from damaged MB has not been changed (compared to the reference frame) these algorithms work well. However, in some cases, the edge direction determination according to boundary pixels of candidate MB can cause low accuracy in MV estimation. When the boundary edges' direction of damaged MB is changed with respect to the reference frame, these algorithms will not lead to acceptable results. Moreover, if the candidate MB is from previous stages of error concealment, the efficiency of the algorithms decreases.

Also, one of the important limitations in conventional error concealment methods (such as BMA, DBM, and DTBMA) is the reduction of accuracy in boundary matching process with decreasing adjacent boundaries' numbers of damaged MB. In these methods, for damaged MBs with more adjacent boundaries is not considered any priority. Therefore, error concealment for the damaged MBs with low adjacent boundaries might be done before the damaged MBs with more adjacent boundaries. If the damaged MBs are neighbor with different adjacent boundaries' numbers, the incorrect estimation of the MV from one of them can reduce the estimation accuracy of other MVs. However, in the proposed algorithm error concealment of damaged MBs in each frame is related to error concealment priority list of the same frame.

At first, the priority list is obtained for error concealment of each frame. Then, the list is updated after error concealment of each damaged MB. Using the priority list, the error concealment is done for the damaged MB with the highest priority (with more adjacent boundaries) sooner than other damaged MBs. Therefore, the priority of neighboring MBs from concealed MB increases. The proposed algorithm leads to more accurate estimation of MVs by considering the more priority for damaged MBs; which have more adjacent boundaries. Moreover, it increases the number of available boundaries for damaged MBs with the numbers of lower boundaries and the damaged MVs of them are estimated more accurately in the next stages of error concealment.

In addition to the use of spatial redundancy from current frame, the proposed algorithm utilizes the temporal redundancy among consecutive frames. According to additional boundaries, the proposed algorithm presents an efficient boundary matching criterion for comparing the candidate MBs. The proposed algorithm uses the proposed boundary matching criterion, just when it is assured about the accuracy and continuity of additional boundaries. Therefore, if additional boundaries are available and have necessary conditions, the proposed algorithm utilizes the proposed boundary matching criterion. This criterion supposes that the damaged MV is similar to the one of adjacent MVs from damaged MB. If the supposition is correct, the MBs with similar MVs belong to the same region in the reference frame. These MBs contain most matches in their common boundaries.

Finally, the proposed algorithm uses the best boundary matching of them for error concealment of damaged MB by selecting one of two explained boundary matching criterion of each candidate MBs' boundary. As the proposed algorithm employs these techniques, the accuracy of the boundary matching process for each boundary of the damaged MB is increased. As such, the estimation of the damaged MV is more accurate.

The proposed algorithm exploits each boundary as a vector. Therefore, the consumed time for exploiting the additional boundaries and calculating the proposed matching distortion criterion in each stage is not significant. According to the results of Table II, the proposed algorithm increases the efficiency of error concealment without any significant increase in computational complexity.

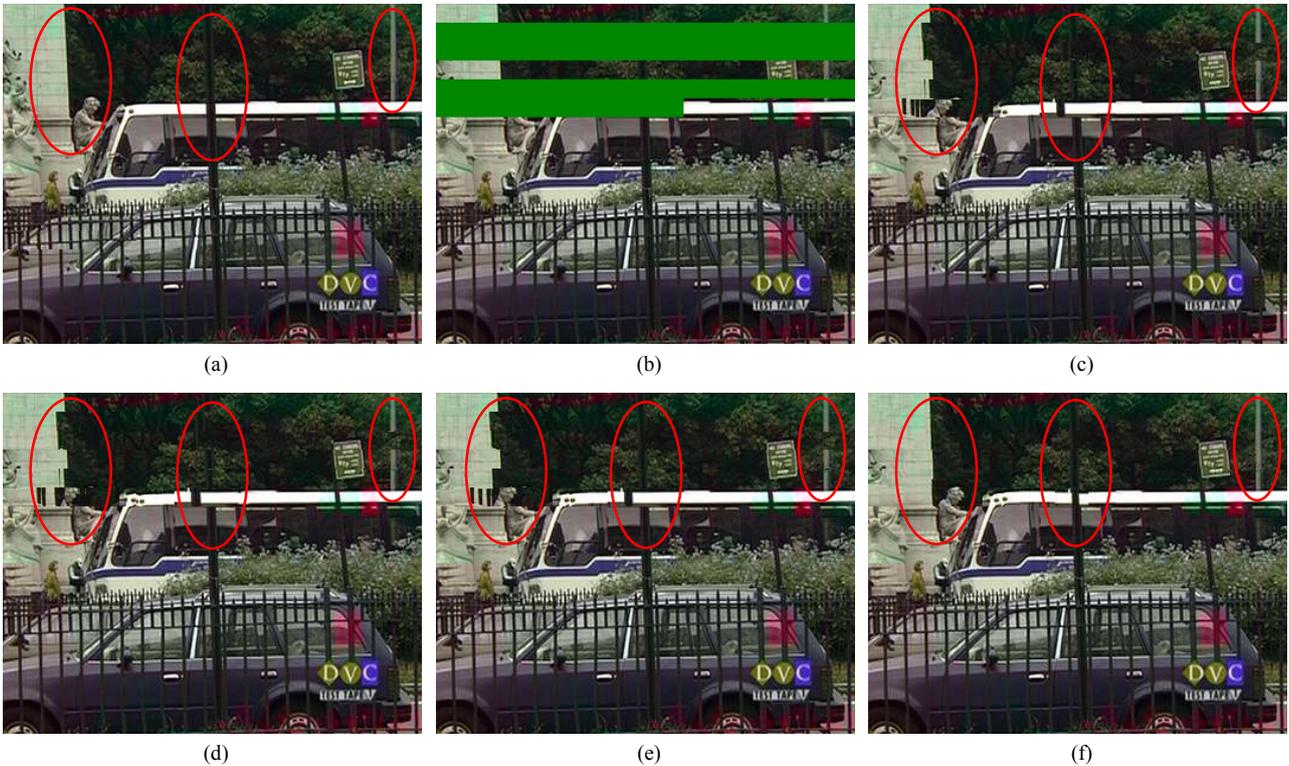

Figure 5.  Subjective video quality comparison for frame 16 in CIF video test sequence "Bus" and average MB missing rate= 20%. (a) Free-error frame  (b) Corrupted frame  (c) Reconstructed frame using BMA  (d) Reconstructed frame using DBM  (e) Reconstructed frame using DTBMA  (f) Reconstructed frame using the proposed algorithm.

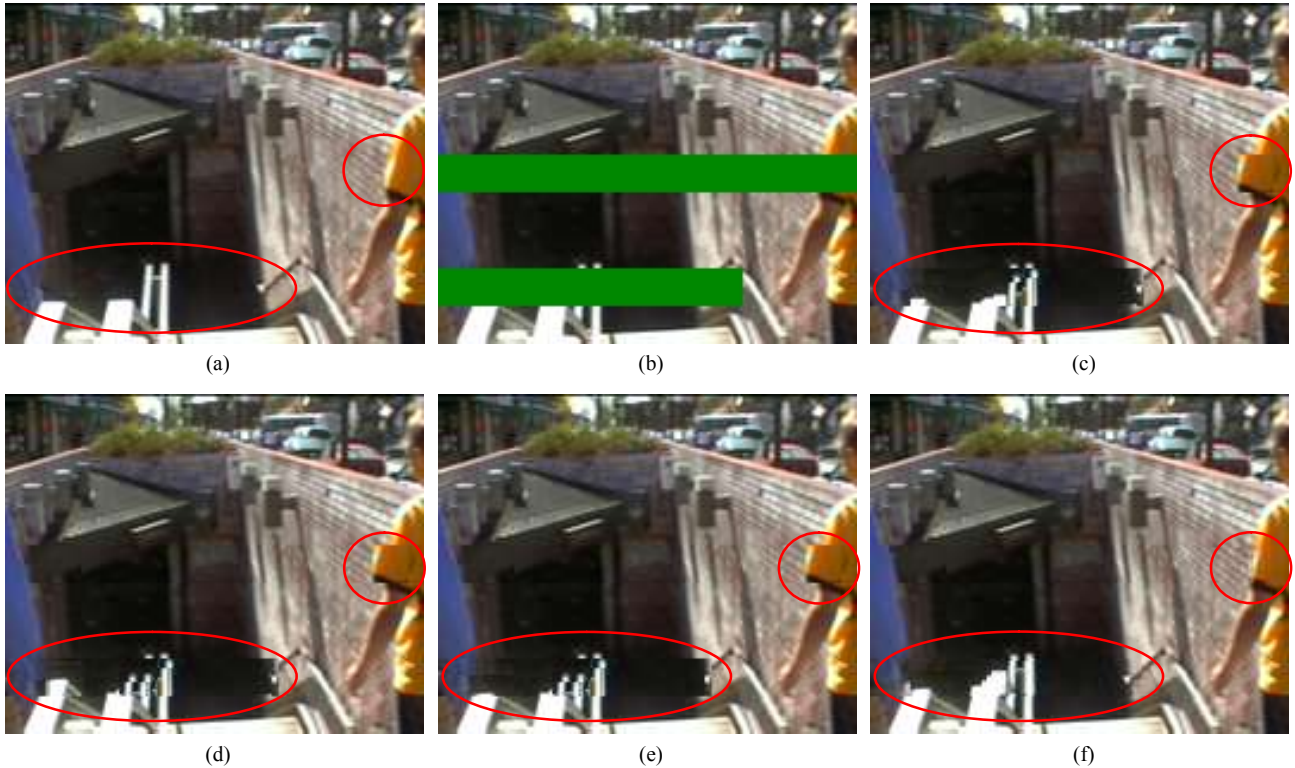

Figure 6. Subjective video quality comparison for frame 31 in QCIF video test sequence "Walk" and average MB missing rate of 20%. (a) Free-error frame (b) corrupted frame (c) reconstructed frame using BMA (d) reconstructed frame using DBM (e) reconstructed frame using DTBMA (f) reconstructed frame using the proposed algorithm.

## V. CONCLUSION

In this paper, an efficient boundary matching algorithm was proposed for more accurate estimation of damaged MVs. The proposed algorithm used the classic boundary matching or the proposed boundary matching criterion to compare the candidate MBs. Also, by identifying the error concealment priority list, damaged MBs are reconstructed according to their priorities. This lead to increase the number of available boundaries for damaged MBs with less boundary and more accurate estimation of their damaged MVs. The proposed algorithm improved the accuracy of MV estimation and decreased the error propagation by using additional data, which their correctness has been analyzed. Experimental results show that the proposed algorithm increases the PSNR for video test sequences in some frames about 4.7, 4.5, and 4.4 dB compared to BMA, DBM, and DTBMA methods, respectively. Also, without considerable computational cost, it improved the objective and subjective quality of reconstructed frames.

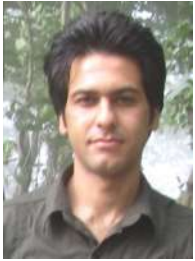
**Seyed Mojtaba Marvasti-Zadeh** was born in Yazd, Iran, on July 15th 1986. He received the Associate's degree in electronics from Technical Faculty of Imam Ali, Yazd, Iran, in 2008. Then, he received the B.Sc. degree in electrical engineering from the Department of Electrical and Electronic Engineering, Science and Arts University, Iran, in 2010. He was awarded as the best graduate student of Electrical and Electronic Departments, Technical Faculty of Imam Ali and Science and Arts University, in 2008 and 2010, respectively. He is now a graduate student of Communications field in the Department of Electrical and Computer Engineering, Yazd University. His research interests include image and video signal processing with a special focus on error concealment of video sequences in error-prone environments.

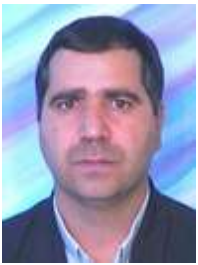
**Hossein Ghanei-Yakhdan** was born in Yazd, Iran in 1966. He received the B.Sc. degree in electrical engineering from Isfahan University of Technology in 1989, the M.Sc. degree in 1993 from K. N. Toosi University of Technology and the Ph.D degree in 2009 from Ferdowsi University of Mashhad. Since 1994, he has been with the Department of Electrical and Computer Engineering, Yazd University. His research interests are in digital video and image processing, error concealment and error-resilient coding for video communication and digital watermarking.

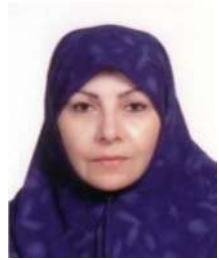
**Shohreh Kasaei** (M'05-SM'07) received her B.Sc. degree from the Department of Electronics, Faculty of Electrical and Computer Engineering, Isfahan University of Technology, Iran, in 1986, her M.Sc. degree from the Graduate School of Engineering, Department of Electrical and Electronic Engineering, University of the Ryukyus, Japan, in 1994, and the Ph.D. degree from Signal Processing Research Centre, School of Electrical and Electronic Systems Engineering, Queensland University of Technology, Australia, in 1998. She joined Sharif University of Technology since 1999, where she is currently a full professor and the director of Image Processing Laboratory (IPL). Her research interests include 3D computer vision, 3D object tracking, human activity recognition, multi-resolution texture analysis, scalable video coding, image retrieval, video indexing, face recognition, hyperspectral change detection, video restoration, and fingerprint authentication.